%Paper: quant-ph/9508002
%From: Domenico Giulini <giulini@ruf.uni-freiburg.de>
%Date: Wed, 2 Aug 1995 19:07:46 +0200 (MET DST)

%%%%%%%%%%%%%%%%%%%%%%%%%%%%% PLAIN TeX INPUT FILE %%%%%%%%%%%%%%%%%%%%%
%%%%%%%%%%%%%%%%%%%%%%%%%%%%%%%%%%%%%%%%%%%%%%%%%%%%%%%%%%%%%%%%%%%%%%%%

%% LAYOUT

\magnification\magstep1
\parskip=\medskipamount
\hsize=6 truein
\vsize=8.2 truein
\hoffset=.2 truein
\voffset=0.4truein
\baselineskip=14pt
\tolerance=500

\font\titlefont=cmbx12
 at 10 truept
\font\authorfont=cmcsc10
\font\addressfont=cmsl10 at 10 truept
\font\smallbf=cmbx10 at 10 truept
 4
     %% Sans serife bold face        USE FOR NUMBER
    %% Sans serife demi bold face       FIELDS

%%% MEMBER BEGINSECTION

\outer\def\beginsection#1\par{\vskip0pt plus.2\vsize\penalty-150
\vskip0pt plus-.2\vsize\vskip1.2truecm\vskip\parskip
\message{#1}\leftline{\bf#1}\nobreak\smallskip\noindent}

%%% MEMBER SUBSECTION

\outer\def\subsection#1\par{\vskip0pt plus.2\vsize\penalty-80
\vskip0pt plus-.2\vsize\vskip0.8truecm\vskip\parskip
\message{#1}\leftline{\it#1}\nobreak\smallskip\noindent}

%%% MEMBER FCOUNT

\newcount\notenumber

\def\note{\advance\notenumber by 1
\footnote{$^{\the \notenumber}$}}      %% upper case without any brackets
%\footnote{$^{\{\the \notenumber\}}$}} %% upper case with curly brackets

%%% MEMBER EIGHTPOINT = FOOTNOTESIZE

\newdimen\itemindent \itemindent=13pt
\def\textindent#1{\parindent=\itemindent\let\par=\resetpar%
\indent\llap{#1\enspace}\ignorespaces}

\let\oldpar=\par
\def\resetpar{\oldpar\parindent=0pt\let\par=\oldpar}

\font\ninerm=cmr9 \font\ninesy=cmsy9
\font\eightrm=cmr8 \font\sixrm=cmr6
\font\eighti=cmmi8 \font\sixi=cmmi6
\font\eightsy=cmsy8 \font\sixsy=cmsy6
\font\eightbf=cmbx8 \font\sixbf=cmbx6
\font\eightit=cmti8
\def\eightpoint{\def\rm{\fam0\eightrm}
  \textfont0=\eightrm \scriptfont0=\sixrm \scriptscriptfont0=\fiverm
  \textfont1=\eighti  \scriptfont1=\sixi  \scriptscriptfont1=\fivei
  \textfont2=\eightsy \scriptfont2=\sixsy \scriptscriptfont2=\fivesy
  \textfont3=\tenex   \scriptfont3=\tenex \scriptscriptfont3=\tenex
  \textfont\itfam=\eightit  \def\it{\fam\itfam\eightit}%
  \textfont\bffam=\eightbf  \scriptfont\bffam=\sixbf
  \scriptscriptfont\bffam=\fivebf  \def\bf{\fam\bffam\eightbf}%
  \normalbaselineskip=9pt
  \setbox\strutbox=\hbox{\vrule height7pt depth2pt width0pt}%
  \let\big=\eightbig \normalbaselines\rm}
\catcode`@=11 %
\def\eightbig#1{{\hbox{$\textfont0=\ninerm\textfont2=\ninesy
  \left#1\vbox to6.5pt{}\right.\n@space$}}}
\def\vfootnote#1{\insert\footins\bgroup\eightpoint
  \interlinepenalty=\interfootnotelinepenalty
  \splittopskip=\ht\strutbox %
  \splitmaxdepth=\dp\strutbox %
  \leftskip=0pt \rightskip=0pt \spaceskip=0pt \xspaceskip=0pt
  \textindent{#1}\footstrut\futurelet\next\fo@t}
\catcode`@=12 %

%%% MEMBER DIAG

%%% MEMBER DIAGS

%% DEFINITIONS

\def\O{\cal O}
\def\l{\langle}
\def\r{\rangle}
\def\L{\cal L}
\def\H{\cal H}
\def\G{\cal G}
\def\Z{\cal Z}
\def\P{\cal P}
\def\b{\bar}
\def\integers{{\bf Z}}
\def\reals{{\bf R}}
\def\Hex{{\cal H}_{\rm ex}}
\def\shalf{\hbox{${\textstyle{1\over 2}}$}}
\def\ihbar{\hbox{${\textstyle{i\over\hbar}}$}}

\font\operator=cmss10 scaled \magstep0
\font\smalloperator=cmss8 scaled \magstep0
\def\op#1{\hbox{\operator#1}}
\def\ops#1{\hbox{\smalloperator#1}}
\hyphenation{Schro-din-ger}

%% REAL STUFF BEGINS

\rightline{Freiburg, THEP-95/15}
\rightline{quant-ph/9508002}
\bigskip
{\baselineskip=24 truept
\titlefont
\centerline{ON GALILEI INVARIANCE IN QUANTUM MECHANICS}
\centerline{AND THE BARGMANN SUPERSELECTION RULE}
}

\vskip 1.1 truecm plus .3 truecm minus .2 truecm

\centerline{\authorfont Domenico Giulini\footnote*{
e-mail: giulini@sun2.ruf.uni-freiburg.de}}
\vskip 2 truemm
{\baselineskip=12truept
\addressfont
\centerline{Fakult\"at f\"ur Physik,
Universit\"at Freiburg}
\centerline{Hermann-Herder Strasse 3, D-79104 Freiburg, Germany}
}
\vskip 1.5 truecm plus .3 truecm minus .2 truecm

\centerline{\smallbf Abstract}
\vskip 1 truemm
{\baselineskip=12truept
\leftskip=3truepc
\rightskip=3truepc
\parindent=0pt

{\eightpoint
We reinvestigate Bargmann's superselection rule for the overall mass
of $n$ particles in ordinary quantum mechanics with Galilei invariant
interaction potential. We point out that in order for mass to define
a superselection rule it should be considered as a dynamical variable.
We present a minimal extension of the original dynamics in which mass
it treated as dynamical variable. Here the classical symmetry group
turns out to be given by an $\reals$-extension of the Galilei group
which formerly appeared only at the quantum level. There is now no
obstruction to implement an action of the classical symmetry group
on Hilbert space. We include some comments of a general nature on formal
derivations of superselection rules without dynamical context.

\par}}

\beginsection{Introduction}

It seems to be a generally accepted text-book wisdom that
non-relativistic quantum mechanics has superselection rules for the
total mass $M$ [1-7]. This means that the superposition of two
states, $\psi_+=\psi_M+\psi_{M'}$, corresponding to different overall
masses, $M$ and $M'$, does not define a pure state. This is sometimes
expressed by saying that such superpositions are forbidden. Formally
this really means that the matrix elements
$\l\psi_M\vert{\O} \vert\psi_{M'}\r$ are zero for all observables
$\O$, which is equivalent to saying that the two density matrices
$\rho_+=\vert\psi_+\r\l\psi_+\vert$ and
$\rho_{\rm mix}=\vert\psi_M\r\l\psi_M\vert +
\vert\psi_{M'}\r\l\psi_{M'}\vert$
define the {\it same} expectation value functional on all observables, i.e.,
$\hbox{tr}(\rho_+\O)=\hbox{tr}(\rho_{\rm mix}\O)$ for all $\O$. In
particular, $\psi_+$ does not define a pure state on the observables.

At this point one must wonder how this statement, which is usually
``derived'' within standard quantum mechanics, should actually be
interpreted within that framework. It obviously refers
to a single system whose set of pure states contains $\psi_M$ and
$\psi_{M'}$. But precisely what is that system? In ordinary quantum
mechanics, the masses are fixed parameters which do not label
different states but rather belong to the specification of
the system. In other words, two $n$-particle systems with different
overall mass are really considered to be {\it different} systems.
In order to regard $\psi_M$ and $\psi_{M'}$ as states of the
{\it same} system, the label $M$ must refer to some dynamical
variable. Mass must therefore be treated dynamically and the
quantum theory should contain a corresponding total mass operator
${\op M}$. That it defines a superselection rule is then equivalent to
saying that ${\op M}$ lies in the centre of the algebra of observables.
But once the total mass becomes a dynamical variable, there is at least
no a~priori reason to restrict the observables to those commuting with
${\op M}$. We thus face the following situation: Standard a priori
derivations within non-relativistic Schr\"odinger theory do not treat
total mass as dynamical variable and hence lack a proper interpretation.
It is true that many texts refer to some ``mass operator'' but, to
our knowledge, a dynamical context is never specified. On the other
hand, if mass is a dynamical variable, there is no a~priori reason
for a mass superselection rule. In order to derive it, one needs
additional inputs which must be different in character from mere
formal consistency
%% begin 2. footnote
conditions\note{In field theory such an additional input is, for example,
given by the principle of locality. There it is the restriction to
(quasi-) local observables that causes the algebra of observables to
acquire a non-trivial centre.}.
%% end 2. footnote
But such a derivation has not yet been
given.

We stress that in principle the specification of a dynamical law is
necessary to find the right implementation of the Galilei group
(or an extension thereof) on state space, since it should be
implemented as a (dynamical) symmetry. Specific properties of the
implementation should therefore not be considered independent of the
dynamical context. We will explain the details of the implementation
in the next section which leads to the precise statement of Galilei
invariance of standard quantum
mechanics\note{Sometimes purely kinematical symmetry groups are
invoked in the definition of quantum mechanical state spaces,
before any dynamical laws are given. See e.g. [8]. This should be
distinguished from the dynamical notion of symmetry used here.}.
This is done not for a free particle [1-6] but in the more
general context of $n$ spinless particles with Galilei-invariant
potential. In particular we learn that it is not the Galilei group
that acts on the Hilbert space but a central extension thereof.
We then recall how this implies the standard argument for the
existence of a superselection rule for overall mass.
In section 2 we present a very simple, minimal generalization of
the classical Hamiltonian system which includes the masses as
dynamical variables. It turns out that here the central extension of
the Galilei group -- formerly only needed to implement the symmetry
group in Hilbert space -- now already appears at the classical level.
In section~3 we discuss the Schr\"odinger equation of this extended
model. There is now no discrepancy between the classical symmetry
group and the symmetry group that acts on the Hilbert space, and hence
no a priori reason from kinematics for a superselection rule. We end
with a brief discussion section.

\beginsection{Section 1}

We denote the Galilei group by $\G$ and parameterize it by
an $SO(3)$ rotation matrix $R$, a boost velocity vector
${\vec v}$, a space translation vector $\vec a$, and a
real-valued time translation $b$. To avoid going into
topological considerations and also to accommodate half-integer
spin we should actually take $SU(2)$ instead of $SO(3)$.
In order to not complicate the notation we can do this
implicitly by regarding $R$ as an $SU(2)$ element whose action
on $R^3$ vectors is via the $SO(3)$ projection.
A group element is
thus denoted by $g=(R,\vec v,\vec a,b)$ and the laws for
multiplication and forming the inverse is given by
$$\eqalignno{
g'g    = &
           (R'\,,\,{\vec v}'\,,\,{\vec a}'\,,\,b')
           (R\,,\,\vec v\,,\, \vec a\,,\, b)              &\cr
       = & (R'R\,,\,{\vec v}'+R'\vec v\,,\, {\vec a}'+
            R'\vec a+{\vec v}'b\,,\, b'+b)                &(1.1)\cr
g^{-1} = & (R\,,\,\vec v\,,\,\vec a\,,\,b)^{-1}
       =   (R^{-1}\,,\, -R^{-1}\vec v\,,\, -R^{-1}
           (\vec a-\vec v b)\,,\,b)\,.                    &(1.2)\cr}
$$
We consider $n$ point-particles of
individual masses $m_i$ interacting via a Galilei invariant potential $V$.
The classical configuration space is $R^{3n}$ coordinatized
by the particle positions $\{{\vec x}_i\}$. On this configuration space
$\G$ acts in the standard way:
$$
(\{{\vec x}_i\}\,,\,t)\mathop{\longrightarrow}^{g} g(\{{\vec x}_i\}\,,\,t)
=(\{R{\vec x}_i+\vec a+\vec vt\}\,,\,t+b)\,.
\eqno{(1.3)}
$$
$V$ is Galilei invariant, if and only if it only depends on the
$\shalf n(n-1)$ distances $r_{ij}:=\vert{\vec x}_i-{\vec x_j}\vert$.
In particular it is time independent. The Schr\"odinger equation reads
$$\eqalignno{
{\op L}\psi:&=\left(i\hbar\partial_t-{\op H}\right)\,\psi=0  &(1.4)\cr
{\op H} &=-\sum_{i=1}^n{\hbar^2\over 2m_i}\Delta_i+
V(\{r_{ij}\})\,,                                             &(1.5)\cr}
$$
where we set $\Delta_i=\vec\nabla_i\cdot\vec\nabla_i$ (no
summation over $i$).
The Hilbert space is given by ${\cal H}=L^2(R^{3n},d^{3n}x)$, with
$d^{3n}x$ the standard Lebesgue measure on $R^{3n}$.
The Schr\"odinger equation is integrated by a one-parameter group
of unitary evolution operators,
$\exp\left({-{i\over\hbar}{\op H}t}\right)$, which provide the
following bijective correspondence between $\H$ and the space of
solutions $\psi(\{{\vec x}_i\},t)$ to the Schr\"odinger equation
$(1.4)$:
$$
\psi(\{{\vec x}_i\},t)\leftrightarrow
\exp\left({\ihbar{\op H}t}\right)
\psi(\{{\vec x}_i\},t)=\psi(\{{\vec x}_i\},t=0)\in\H\,.
\eqno{(1.6)}
$$

We now try to consider the Galilei group as symmetry group in the
quantum theory. A priori it is not obvious how an element
$g\in\G$ should act on $\H$. It is important to note that we do not
just wish to find any unitary action, of which there are clearly many,
but rather the particular action that corresponds to a symmetry for
the dynamical equation $(1.4)$. The idea is to use the identification
$(1.6)$ and first determine $g$'s action on the solutions of
${\op L}\psi=0$. These are functions of the coordinates
$(\{{\vec x}_i\},t)$ on which
$g$'s action is given by $(1.3)$. An obvious choice would therefore
consist in shifting the function $\psi$ along $g$ by taking the
composite function $\psi\circ g^{-1}$, just like for a classical field.
But one may easily check that the shifted function does not solve
the Schr\"odinger equation anymore. This can be remedied by also
multiplying $\psi$ with a phase-function $\exp(if_g)$ which we now
determine. We set
$$
T:\,\psi\mathop{\longrightarrow}^{g} {\op T}_g\psi
    = \exp(if_g)\,(\psi\circ g^{-1})
\eqno{(1.7)}
$$
and impose the requirement that the resulting function again solves the
Schr\"odinger equation. Acting with ${\op L}$ on ${\op T}_g\psi$ one
finds
$$\eqalign{
{\op L}({\op T}_g\psi)=
& \exp(if_g)\,({\op L}\psi\circ g^{-1}) \cr
& +i\hbar\exp(if_g)\,\sum_{i=1}^n\left({\hbar\over m_i}\vec\nabla_i f_g
  -\vec v\right)\cdot R\cdot\left(\vec\nabla_i\psi\circ g^{-1}\right)\cr
& -\hbar\exp(if_g)\,\left(\partial_tf_g+\sum_{i=1}^n{\hbar\over 2m_i}
   (\vec\nabla_i f_g)^2-i\sum_{i=1}^n{\hbar\over 2m_i}\Delta_i f_g\right)\,
    \left(\psi\circ g^{-1}\right)\,.\cr}
\eqno{(1.8)}
$$
This shows that ${\op T}_g$ transforms solutions of the Schr\"odinger
equation to solutions, if and only if the extra two terms in $(1.8)$
vanish. This is equivalent to
$$
f_g(\{{\vec x}_i\},t)
=    {M\over\hbar}\left(\vec v\cdot\vec R-
     \shalf{\vec v}^2t+c_g\right)\,,
\eqno{(1.9)}
$$
where $M=\sum_{i}m_i$ is the total mass and
$\vec R={1\over M}\sum_im_i{\vec x}_i$ the centre-of-mass-vector.
$c_g$ is a $g$-dependent integration constant. A convenient
choice is $c_g={1\over 2}{\vec v}^2b-\vec v\cdot\vec a$,
which leaves us with the transformation law
$$
({\op T}_g\psi)(\{{\vec x}_a\},t)=\exp\left\{\ihbar M
\left[\vec v\cdot(\vec R-\vec a)-\shalf{\vec v}^2(t-b)\right]\right\}\
\psi(g^{-1}(\{{\vec x}_i\},t))\,.
\eqno{(1.10)}
$$
The fact that ${\op T}_g\psi$ satisfies the Schr\"odinger
%% begin footnote
equation\note{To ease comparison with similar formulae in the
literature we remark that the transformation law $(1.10)$ may
alternatively be written in the form
${\ops T}_g\psi=(\exp(i{\tilde{f}}_g)\,\psi)\circ g^{-1}$,
where ${\tilde{f}}_g=f_g\circ g$. Using expression $(1.9)$
with our choice of $c_g$, this leads to ${\tilde{f}}_g(\{{\vec x}_a\},t)
=\vec v\cdot R\cdot\vec R+{1\over 2}{\vec v}^2t$.}
%% end footnote
can be equivalently expressed by
$({\op T}_g\psi)(t)=\exp(-\ihbar{\op H}t)(({\op T}_g\psi)(t=0))$, which means
that we just need to put $t=0$ in (1.10) in order to obtain $g$'s
action on $\H$, which we call $U:\, g\rightarrow {\op U}_g$,
$$\eqalignno{
{\op U}_g\psi(\{{\vec x}_i\})
= & \exp\left\{\ihbar M
  \left[(\vec v\cdot(\vec R-\vec a)+\shalf{\vec v}^2\,b\right]\right\}
&\cr
\times & \left(\exp(\ihbar{\op H}b)\psi\right)
         (\{R^{-1}({\vec x}_i-\vec a+\vec vb)\})  \,.
&(1.11)\cr}
$$
This action is quite obviously unitary.

However, having found the transformation law for each $g\in \G$
does not imply a representation of $\G$ on $\H$. In fact one now
finds a phase difference between the Galilei transformation
${\op U}_{g'g}$ and the composite transformation
${\op U}_{g'}{\op U}_g$:
$$\eqalignno{
  {\op U}_{g'}{\op U}_g
&=\exp\left(\ihbar M\,\xi(g',g)\right)\,{\op U}_{g'g}
&  (1.12) \cr
  \xi(g',g)
&=\vec {v'}\cdot R'\cdot\vec a+\shalf{\vec {v'}}^2b\,.
&  (1.13)\cr}
$$
It is straightforward to check the condition
$$
\delta\xi(g'',g',g):=\xi(g''g',g)-\xi(g'',g'g)+\xi(g'',g')-\xi(g',g)=0
\eqno{(1.14)}
$$
which implies associativity of the multiplication law $(1.12)$.
Had we chosen different constants ${c'}_g=c_g+\delta_g$ in $(1.9)$,
$\xi$ would be redefined according to
$$
\xi(g'g)\mapsto\xi'(g',g)
=\xi(g',g)-\delta_{g'g}+\delta_{g'}+\delta_g\,,
\eqno{(1.15)}
$$
which also satisfies $(1.14)$. But the crucial observation is that
the phases $\xi$ cannot be made to zero by such a redefinition. Indeed,
if such a choice existed, it would at the same time remove all phase
factors on subgroups. That this is not possible can be easily seen
by restricting to the abelian subgroup of translations and boosts
on which the induced multiplier phases are
$
\xi(({\vec v}',{\vec a}'), ({\vec v},{\vec a}))={\vec v}'\cdot\vec a
$.
But on abelian groups a redefinition of the form $(1.15)$ is necessarily
symmetric in the group elements $g'$ and $g$ and cannot possibly remove
a non-symmetric expression.

Due to the unavoidable presence of the phase factors in the
transformation law $(1.11)$ we do not have a representation of the
Galilei group on our Hilbert space, but rather a projective
representation. Equivalently [9], we can have a proper
representation of a slightly larger group, the so-called extended
Galilei group $\b{\G}$, which is a central extension of $\G$ by
a group $\Z$ isomorphic to
%% begin footnote
$\reals$\note{As is e.g. shown in [9], ray representations are in
bijective correspondence to central $U(1)$-extensions. Here we
consider $\reals$-extensions which form the universal cover.
This becomes significant in section 2.}
%% end footnote
This means that all of $\Z$
lies in the centre of $\b{\G}$ and that the quotient group $\b{\G}/\Z$
is isomorphic to $\G$. A group element $\b g\in\b{\G}$ is now
written in the form $\b g=(\theta,g)$, where $\theta\in\Z$ and
$g\in\G$ as in $(1.1)$. The multiplication law is given by
$$
{\b g}'\b g=(\theta',g')(\theta,g)=(\theta'+\theta+\xi(g',g)\,,\,g'g)
\eqno{(1.16)}
$$
which serves to define the following unitary representation $\b U$
of $\b{\G}$ on $\H$:
$$\eqalignno{
& {\b {\op U}}_{\b g}\psi=
  \exp\left(\ihbar M\theta\right)\,{\op U}_g\psi\,,
& (1.17)\cr
& {\b {\op U}}_{{\b g}'}{\b {\op U}}_{\b g}=
  {\b {\op U}}_{{\b g}'\b g}\,,
& (1.18)\cr}
$$
with ${\op U}_g\psi$ as in $(1.11)$.

 From $(1.11)$ and $(1.17)$ we infer that an infinitesimal transformation
${\b {\op U}}_{\delta\b g}$ with infinitesimal parameters
$\delta R_{ab}=\varepsilon_{abc}\delta k_b$, $\delta\vec v$,
$\delta\vec a$, $\delta b$, and $\delta\theta$ is given by
$$
{\b {\op U}}_{\delta\b g}
=\delta\vec k\cdot\vec{\op D}+\delta\vec v\cdot\vec{\op V}
 +\delta\vec a\cdot\vec {\op A}+\delta b\,{\op B}
 +\delta\theta\,{\op Z} \,,
\eqno{(1.19)}
$$
where
$$\eqalignno{
\vec{\op D}:=& -\sum_{i=1}^n(\vec x_i\times\vec\nabla_i)  &(1.20a)\cr
\vec{\op V}:=& \ihbar M \vec R                            &(1.20b)\cr
\vec{\op A}:=& -\sum_{i=1}^n \vec\nabla_i                 &(1.20c)\cr
    {\op B}:=& \ihbar {\op H}                             &(1.20d)\cr
    {\op Z}:=& \ihbar M {\op I}\,.                        &(1.20e)\cr}
$$
The symbol ${\op I}$ denotes the identity operator.
The generators $(1.20)$ satisfy the Lie algebra relations for the
group $\b{\G}$, whose non-vanishing commutators are given by
$$\eqalignno{
& [{\op D}_a,{\op D}_b]=\varepsilon_{abc}{\op D}_c        &(1.21a)\cr
& [{\op D}_a,{\op V}_b]=\varepsilon_{abc}{\op V}_c        &(1.21b)\cr
& [{\op D}_a,{\op A}_b]=\varepsilon_{abc}{\op A}_c        &(1.21c)\cr
& [{\op V}_a,{\op A}_b]=\delta_{ab}{\op Z}                &(1.21d)\cr
& [{\op V}_a,{\op B}]  ={\op A}_a  \,.                    &(1.21e)\cr}
$$
We call this Lie algebra $\b{\L}$. It differs from the corresponding
one for the Galilei group only by $(1.21d)$ whose right hand side is
now proportional to the central element ${\op Z}$ instead of being
zero. Besides ${\op Z}$ there are two more central elements in the
enveloping algebra of $\b{\L}$ (i.e. Casimir elements):
$$\eqalignno{
{\vec{\op S}}^2
&=({\op Z}\vec{\op D}-\vec{\op V}\times\vec{\op A})^2
& (1.22)\cr
{\op K}
&={\vec{\op A}}^2-2{\op Z}{\op B}\,.
&(1.23)\cr}
$$
In the representation $(1.20)$ the vector operator $\vec{\op S}$
has the interpretation of $M/\hbar^2$ times the internal angular
%% begin footnote
momentum\note{The term `spin' is already reserved for the possible
internal angular momentum of each particle. We shall not use this
term for the translation-invariant part of angular momentum.}
%% end footnote
and ${\op K}$ corresponds to $2M/\hbar^2$ times the
internal energy, i.e. the total energy minus the kinetic energy
of the center of mass motion. This interpretation may be taken over
to any irreducible representation corresponding to strictly
positive eigenvalues of the Casimir element ${\op Z}$.
In view of $(1.20e)$, ${\op Z}$ is sometimes
given the interpretation of $\ihbar{\op M}$ with ${\op M}$ as
operator for the overall mass. But since mass is none of our
dynamical variables this does, in our opinion, not really make
much sense in the present context.

The superselection rule first stated by Bargmann [1] is usually
motivated in the following manner: Let us restrict to the
subgroup generated by space translations and boosts and let
$g=(1,0,\vec a,0)$ and $g'=(1,\vec v,0,0)$ be the group elements
for a spatial translation and a boost respectively. On one hand
we have $g'^{-1}g^{-1}g'g=1$, whereas the corresponding operators
on the Hilbert space, ${\op U}_{g'}$ and ${\op U}_{g}$, obey
${\op U}^{-1}_{g'}{\op U}^{-1}_g{\op U}_{g'}{\op U}_g=
\exp(-\ihbar M \vec v'\cdot\vec a){\op I}$ (compare $(1.12-13)$).
Applying this combination to a superposition $\psi_M+\psi_{M'}$
results in a relative phase factor of $\exp(\ihbar\vec v'\vec a(M-M'))$
and therefore a different ray, unless $M=M'$. But the identity
Galilei transformation should not alter physical states. Hence
$\psi_M+\psi_{M'}$ cannot represent a physical state if
$M\not =M'$. Let us describe this situation in a slightly more
geometric fashion. We are given two Hilbert spaces, ${\H}_M$
and ${\H}_{M'}$, and their associated projective spaces of
rays, ${\P\H}_M$ and ${\P\H}_{M'}$. That each Hilbert space
carries a projective representation of $\G$ means that $\G$
acts on ${\P\H}_M$ and ${\P\H}_{M'}$. But, as just shown,
$\G$ does not act on ${\P}({\H}_M\oplus {\H}_{M'})$. The largest
subset it acts on is the disjoint union
${\P\H}_M\cup{\P\H}_{M'}$. Elements in the complement cannot
belong to the set of pure states which, by assumption, admits an
action of the Galilei group.

Let us now look at this situation from the extended
Galilei group $\b{\G}$. The left hand side of $(2.21d)$ just
represents the infinitesimal version of the combination of
translations and boosts given above. The requirement that physical
states are unchanged by this combination is then equivalent to the
requirement that physical states are
%%%% begin footnote
eigenstates\note{We shall generally refer to generalized eigenstates,
which occur for operators with continuous spectra, also simply as
eigenstates. The appropriate constructions -- well known from standard
quantum mechanics -- are implicitly understood at these points.}
%%%%% end footnote
of ${\op M}=-i\hbar{\op Z}$. Geometrically the situation is now
different. We now have proper actions of $\b{\G}$ on all the
Hilbert spaces and hence on all corresponding projective spaces of
rays. However, whereas $\Z$ acts trivially on ${\P\H}_M$ and
${\P\H}_{M'}$, it acts non-trivially on ${\P}({\H}_M\oplus{\H}_{M'})$.
The superselection principle now declares that pure states must be
fixed-points of $\Z$'s action, which for ${\P}({\H}_M\oplus{\H}_{M'})$
are given by the set ${\P\H}_M\cup{\P\H}_{M'}$.

We have argued at the beginning that in order to make sense of a
mass superselection rule one should regard mass as dynamical
variable. In the quantum theory, its associated self-adjoint
operator, ${\op M}$, then generates a one-parameter group of
unitary transformations which we may identify with $\Z$.
Any theory with dynamical mass is therefore expected to
admit $\b{\G}$ rather than $\G$ as symmetry group. In the next
section we show how this can arise at the classical level.

\beginsection{Section 2}

In this section we minimally extend the Hamiltonian system
considered so far in order to also treat the masses $\{m_i\}$
in a dynamical fashion. The idea is really simple, namely to
just retain the original Hamiltonian function,
$$
H=\sum_{i=1}^n {{\vec p}^2_i\over 2m_i}+V(\{\vec x_i\},\{m_i\})\,,
\eqno{(2.1)}
$$
and adjoin the $2n$ real-valued canonical variables
$\{(\zeta_i,m_i)\}$, $i=1,..,n$. We consider $\{\zeta_i\}$ as
new coordinates and $\{m_i\}$ the corresponding momenta. They satisfy
the standard canonical commutation
%% begin footnote
relations\note{We allow $m_i$ to take values on the whole real axis.
If one restricted $m_i$ to the positive real axis, the procedure would
be different at this point.}
%% end footnote
$\{\zeta_i,m_j\}=\delta_{ij}$. Correspondingly, we have to add a
kinetic term $\sum_i m_i{\dot\zeta}_i$ to the action:
$$
S=\int dt\left\{\sum_{i=1}^n\left({\vec p}_i {\dot{\vec x}}_i
   +m_i{\dot\zeta}_i\right)-\sum_{i=1}^n{{\vec p}_i\over 2 m_i}
   -V(\{{\vec x}_i\},\{m_i\})\right\}\,.
\eqno{(2.2)}
$$
Variations with respect to ${\vec x}_i$, ${\vec p}_i$, $\zeta_i$, and
$m_i$ yield the Hamiltonian equations:
$$\eqalignno{
{\dot{\vec p}}_i & = -\vec\nabla_{x_i}V                  &(2.3a)\cr
{\dot{\vec x}}_i & = {{\vec p}_i\over m_i}               &(2.3b)\cr
{\dot m}_i       & = 0                                   &(2.3c)\cr
{\dot\zeta}_i      & = \nabla_{m_i}V-{{\vec p}_i^2
                     \over 2m_i^2}\,.                    &(2.3d)\cr}
$$
Equations $(2.3c)$ just imply conservation of the individual masses.
Inserting constant $m_i$ into $(2.3a-b)$ yields the standard equations
which also allow to derive the law of energy conservation in its
familiar form: $\sum_i\shalf m_i{\dot{\vec x}}_i^2+V=$ const.
Having found solutions $(\{{\vec x}_i(t)\},\{{\vec p}_i(t)\})$ we can
easily integrate $(2.3d)$:
$$
\zeta(t)=\int^t dt'\ \left\{\nabla_{m_i}V(\{{\vec x}_i(t')\},\{m_i\})
       -\shalf{\vec x}_i^2(t')\right\}  \,.
\eqno{(2.4)}
$$

Next we wish to investigate the invariance properties of the given
dynamics under the Galilei group. We assume invariance of the masses
$m_i$ which obviously implies the invariance of $(2.3a-c)$ under
general Galilei transformations. To investigate invariance of
$(2.3d)$, we write down this equation for the transformed solution
curves, insert ${\vec {p'}}_i(t)=R{\vec p}_i(t)+m_i\vec v$, and
subtract $(2.3d)$ for the untransformed solution curves.
This yields to
$$\eqalignno{
{\dot\zeta}'_i(t) & =
      {\dot\zeta}_i(t)-\vec v\cdot R\cdot{\dot{\vec x}}_i(t)
      +\shalf\vec v^2t &\cr
\hbox{or}\quad
{\zeta'}_i(t) & =\zeta_i(t)-\vec v\cdot R\cdot{\vec x}_i(t)-
             \shalf\vec v^2t+\gamma_g\,,            &(2.5)\cr}
$$
where $\gamma_g$ is some constant. We will set it to zero.

The transformation law $(2.5)$ does not define an action of the Galilei
group on phase space. However, it defines an action of the extended
group $\b{\G}$. To see this in more detail, we explicitly display the
transformation law on configuration
%% begin footnote
space\note{Since the transformation law for the new momenta $\{m_i\}$
is trivial we may restrict attention to the configuration space.}
%% end footnote
(including the time axis) for the group element
$\b g=(\theta,R,\vec v,\vec a,b)\in\b{\G}$:
$$
\b g\,\left(\{{\vec x}_i\},\{\zeta_i\},t\right)=
\left(\{R{\vec x}_i+\vec v t+\vec a\}\,,\,
\{\zeta_i-(\theta+\vec v\cdot R\cdot{\vec x}_i+\shalf\vec v^2t)\}\,,\,
t+b\right)\,.
\eqno{(2.6)}
$$
It is now straightforward to verify that a transformation
$\b g=(\theta,g)$ followed by a transformation $\b g'=(\theta',g')$
equals a transformation $\b g'\b g=(\theta'+\theta+\xi(g',g)\,,\,g'g)$,
where $\xi(g',g)=\vec v'\cdot R'\vec a+\shalf\vec v^2b$, as in $(1.13)$.
Relation $(1.14)$ then ensures associativity so that $(2.6)$ defines
indeed an action of the group $\b{\G}$ on configuration- and phase space
(including the time axis). The inverse element to $(\theta,g)$ is also
easily calculated:
$$
\left(\theta,R,\vec v,\vec a,b\right)^{-1}=
\left(-\theta+\vec v\cdot\vec a-\shalf\vec v^2b\,,\,
R^{-1}\,,\, -R^{-1}\cdot\vec v\,,\,
-R^{-1}\cdot(\vec a-\vec vb)\,,\, -b\right).
\eqno{(2.7)}
$$
Hence for its action on configuration space and time axis:
$$\eqalignno{
&(\theta,g)^{-1}(\{{\vec x}_i\},\{\zeta_i\},t)=&\cr
&\left(\{R^{-1}\cdot({\vec x}_i-\vec v (t-b)-\vec a\}\,,\,
\{\zeta_i+\theta+\vec v\cdot({\vec x}_i-\vec a)-\shalf\vec v^2(t-b)\}
\,,\, t-b\right).
&(2.8)\cr}
$$

\beginsection{Section 3}

In the last section we have seen how the extended Galilei group $\b{\G}$
arises as the symmetry group on the classical level if the masses are
treated as dynamical variables. In this section we show that the same
group acts as symmetries in the corresponding quantum mechanical theory.
There is therefore no need for a superselection rule in this model.

In the quantum mechanical treatment of this model the Hilbert space
is now given by $\Hex=L^2(R^{4n},d^{3n}x\, d^{n}\zeta)$,
where $R^{4n}$ is spanned by the coordinates $\{{\vec x}_i\}$ and
$\{\zeta_i\}$ and the measure is just the product of the Lebesgue measures.
Since the Hamiltonian $(2.1)$ does not depend on the coordinates
$\{\zeta_i\}$ it is convenient to first perform a Fourier transform in the
$\zeta_i-m_i$ variables:
$$
\Psi(\{{\vec x}_i\},\{\zeta_i\},t)=(2\pi\hbar)^{-{n\over 2}}
\int d^n m\, \exp\left({i\over\hbar}\sum_{i=1}^n m_i\zeta_i\right)\
\Phi(\{{\vec x}_i\},\{m_i\},t)\,.
\eqno{(3.1)}
$$
The Schr\"odinger equation for $\Psi$ is then equivalent to
$$
i\hbar\partial_t\Phi(\{{\vec x}_i\},\{m_i\},t) =
\left\{-\sum_{i=1}^n {\hbar^2\over 2m_i}\Delta_i +
V(\{{\vec x}_i\},\{m_i\})\right\}\, \Phi(\{{\vec x}_i\},\{m_i\},t)\,.
\eqno{(3.2)}
$$
Using the Fourier isomorphism we may think of $\Hex$ as a direct
integral of Hilbert spaces ${\H}_{\{m_i\}}$ each of which is
isomorphic to ${\H}=L^2(R^{3n},d^3x)$:
$$
\Hex=\int_{{\reals}^n} d^nm\,{\H}_{\{m_i\}}\,,
\eqno{(3.3)}
$$
and where a vector $\Phi$ is considered as a map
$\Phi: R^n\rightarrow \H$, $\{m_i\}\mapsto \Phi_{\{m_i\}}$ with
$\Phi_{\{m_i\}}(\{{\vec x}_i\})=\Phi(\{{\vec x}_i\},\{m_i\})$.
It is not difficult to show the invariance of the Schr\"odinger equation
under the group $\b{\G}$. Its action on solutions
$\Psi(\{{\vec x}_i\},\{\zeta_i\},t)$ is now simply defined by
composition:
$$
{\b {\op T}}_{\b g}\Psi:=\Psi\circ {\b g}^{-1}
\eqno{(3.4)}
$$
where the action of ${\b g}^{-1}$ on $(\{{\vec x}_i\},\{\zeta_i\},t)$
is given by $(2.8)$. Using $(3.1)$ this is seen to induce
transformations ${\op T}^{\{m_i\}}_{\b g}$ for the $\Phi_{\{m_i\}}$:
$$ \eqalignno{
{\b {\op T}}^{\{m_i\}}_{\b g}\Phi_{\{m_i\}}(\{{\vec x}_i\},t)
= &
  \exp\left\{\ihbar M \left[\theta+\vec v\cdot(\vec R-\vec a)-
  \shalf\vec v^2(t-b)\right]\right\}                   &\cr
  \times & \Phi_{\{m_i\}}(g^{-1}(\{{\vec x}_i\},t))\,. &(3.5)\cr}
$$
Comparing this to $(1.10)$ and $(1.17)$ we see that this is just the
action of $\b{\G}$ on the solution space of the Schr\"odinger equation
found in section 1. For each set $\{m_i\}$ we now have a representation,
${\b U}^{\{m_i\}}:g\rightarrow{\op U}^{\{m_i\}}$, given by (compare
$(1.11)$):
$$\eqalignno{
{\b {\op U}}^{\{m_i\}}_{\b g}\Phi_{\{m_i\}}(\{{\vec x}_i\})
= & \exp\left\{\ihbar M
  \left[(\vec v\cdot(\vec R-\vec a)+\shalf{\vec v}^2\,b\right]\right\}
&\cr
\times & \left(\exp(\ihbar{\op H}b)\Phi_{\{m_i\}}\right)
         (\{R^{-1}({\vec x}_i-\vec a+\vec vb)\})
&(3.6)\cr}
$$
where $M$, $\vec R$, and ${\op H}$ should be understood as functions of
the mass variables $\{m_i\}$. Hence, the representation of $\b{\G}$ on $\Hex$
can be written as a direct integral of representations
$$
\b U=\int_{\reals^n}d^nm\,{\b U}^{\{m_i\}}\,.
\eqno{(3.7)}
$$
The representation ${\b U}^{\{m_i\}}$, restricted to the central
subgroup $\Z$ (isomorphic to \reals\ and generated by ${\op Z}$) has a
kernel which is given by $\{{\hbar\over M}2\pi q,\,q\in \integers\}$,
where $\integers$ denotes the integers. This defines a different
subgroup of $\Z$ for different $M$, thereby showing again that
representations for different $M$ are inequivalent. The requirement
on observables to commute with the action of $\Z$, which forms
the group of translations along the ``diagonal'' in $\zeta$-space,
then gives rise to the continuous superselection
rule\note{For some background material on continuous superselection
rules, see e.g. [10][11].} for the overall mass.  But this
superselection rule is by no means necessary in order to implement
an action of the classical symmetry group on quantum mechanical state
space. The price to pay is to recognize $\G$ rather than $\G$ as
classical symmetry group.

\beginsection{Discussion}

The transformation $(3.1)$ should be considered as expansion in
common eigenstates of the operators $-i\hbar\nabla_{\zeta_i}$ which
generate translations in $\zeta_i$ and correspond to the operators
for the individual mass $m_i$. The Schr\"odinger equation allows
to separate the $\zeta_i$ motions, just like the center of
mass motion is separable in standard translation invariant problems.
Expanding in plane waves for these ignorable coordinates leads to a
reduced equation which in our case is just the ordinary Schr\"odinger
equation for fixed masses. It is indeed instructive to compare the
situation to ordinary quantum mechanics in a translation invariant
context. In the latter case, translation invariance is not
interpreted to generally prevent us from forming superpositions of
plane waves which correspond to quasi localized wave packets.
Clearly, in order to prepare such states we need to break translation
invariance. The resulting states are then not momentum eigenstates
and to manufacture them we need operators which do not commute with
translations. We usually do not regard this as a difficulty. Quite the
contrary, in order to view translation invariance as a proper
physical symmetry we have to regard the translated states as equally
valid but decidably different states. This is what distinguishes a
symmetry from a mere redundancy. Redundancies are described by gauge
symmetries which are conceptually different from physical symmetries
and also lead to different mathematical consequences. Regarding
translations as gauge symmetries is equivalent to saying that motions
in the translational directions do not change physical states.
But in our example, translating the $\zeta_i$'s means to change the
physical state. For example, shifting the system in real time along
the $\zeta_i$'s costs action, according to $(2.2)$. This would not be
the case if we were considering pure gauge degrees of freedom. Within
the framework of our dynamical model we thus talk about physically
existing degrees of freedom. Stating a superselection rules for the
masses must therefore be equivalent to stating that for some
{\it physical} reason we cannot localize the system in $\zeta_i$-space.
It seems plausible that many derivations of superselection
rules from purely formal arguments in fact make at least one
contingent physical assumption of that sort. For better
understanding the actual physical input one should in our opinion
1.) find the right dynamical theory in which the relevant
quantities are manifestly dynamical and 2.) address the question
of what is actually measurable within that framework.
Similar views were also expressed in [12].
The present model with dynamical masses is also meant to
illustrate this point of view.

\vskip1.0truecm
\centerline{\bf ACKNOWLEDGEMENTS}
\medskip
I thank K. Kucha\v r and J. Kupsch for valuable discussions.

%%%%%%%%%%%%%%%%%%%%%%%%%%%%%%%%%%%%%%%%%%%%%%%%%%%%%%%%%%%

%%%%%%%%%%%%%%%%%%%%%%% REFERENCES %%%%%%%%%%%%%%%%%%%%%%%%

\vfill\eject
\beginsection References

{\parskip=0.1truecm

\item{[1]}  V. Bargmann, On Unitary Ray Representations of Continuous
          Groups. {\it Ann. of Math.}, {\bf 59} (1954), 1-46

\item{[2]}  J-M. Levy-Leblond, Galilei Group and Nonrelativistic
          Quantum Mechanics. {\it Jour. Math. Phys.}, {\bf 4}
          (1963), 776-788

\item{[3]}  F.A. Kaempffer, Concepts in Quantum Mechanics, App. 7.
          Academeic Press, New York - London 1965

\item{[4]}  H. Primas and U. M\"uller-Herold, Elementare Quantenchemie,
          \S 3.4.4. Verlag B.G. Teubner, Stuttgart 1984

\item{[5]}  H. Primas, Chemistry, Quantum Mechanics and Reductionism,
          \S3.3, second corrected edition. Springer Verlag,
          Berlin - Heidelberg 1983

\item{[6]}  A. Galindo and P. Pascual, Quantum Mechanics I, \S 7.14.
           Springer Verlag, Berlin - Heidelberg 1990

\item{[7]} A.S. Wightman, Superselection Rules; Old and New.
           {\it Nouvo Cimento}, {\bf 110 B} (1995), 751-769

\item{[8]}  P.P. Divakaran, Symmetries and Quantization: Structure
          of the State Space. {\it Rev. Math. Phys.}, {\bf 6}
          (1994), 167-205

\item{[9]}  M.S. Raghunathan, Universal Central Extensions.
          {\it Rev. Math. Phys.}, {\bf 6} (1994), 207-225

\item{[10]} C. Piron, Les R\`egles de Supers\'election Continues.
            {\it Helv. Phys. Acta}, {\bf 42} (1969), 330-338

\item{[11]}  H. Araki, A Remark on Machida-Namiki Theory of
             Measurement.
             {\it Prog. Theor. Phys.}, {\bf 64} (1980), 719-730

\item{[12]}  D. Giulini and C. Kiefer and H.D. Zeh, Symmetries,
          Superselection Rules, and Decoherence. {\it Phys. Lett. A},
          {\bf 199} (1995), 291-298

}
\end